\documentclass{ws-p9-75x6-50}

\begin{document}

\title{CMB anisotropy from spatial correlations of clusters of 
galaxies}

\author{Eiichiro Komatsu$^{1,2,3}$, Tetsu Kitayama$^4$,
Alexandre Refregier$^5$, David N. Spergel$^{1,6}$ and Ue-Li Pen$^7$}

\address{
$^1$ Department of Astrophysical Sciences, Princeton University, 
Princeton, NJ 08544, USA\\
$^2$ School of Natural Sciences, Institute for Advanced Study, 
Princeton, NJ 08540, USA\\
$^3$ Astronomical Institute, T\^ohoku University, 
Aoba, Sendai 980-8578, Japan\\
$^4$ Department of Physics, Tokyo Metropolitan University, 
Hachioji, Tokyo, 192-0397\\
$^5$ Institute of Astronomy, Madingley Road, University of Cambridge,
Cambridge CB3 OHA, England\\
$^6$ Keck Distinguished Visiting Professor,
School of Natural Sciences, Institute for Advanced Study, 
Princeton, NJ 08540, USA\\
$^7$ Canadian Institute of Theoretical Astrophysics, 
University of Toronto, 60 St. George St., Toronto, Canada\\
E-mail: komatsu@astro.princeton.edu}


\maketitle

\abstracts{The Sunyaev-Zel'dovich (SZ) effect from clusters of
galaxies is a dominant source of secondary cosmic microwave background
(CMB) anisotropy in the low-redshift universe.  We present analytic
predictions for the CMB power spectrum from massive halos arising from
the SZ effect.  Since halos are discrete, the power spectrum consists
of a Poisson and a correlation term. The latter is always smaller than
the former, which is entirely dominated by nearby bright massive
halos, i.e., by rich clusters.  In practice however, those bright
clusters are easy to indentify and can thus be subtracted from the
map.  After this subtraction, the correlation term dominates
degree-scale fluctuations over the Poisson term, as the main
contribution to the correlation term comes from distant clusters.  We
compare the signal of the correlation term to the expected sensitivity
for the Planck experiment for the SZ effect, and find that the
correlation term is detectable.  Since the degree scale spectrum is
quite insensitive to the highly uncertain core structures of halos,
our predictions are robust on these scales.  Measuring the correlation
term on degree scales thus cleanly probes the clustering of distant
halos. This has not been measured yet, mainly because optical and
X-ray surveys are not sufficiently sensitive to include such distant
clusters and groups. Our analytic predictions are also compared to
adiabatic hydrodynamic simulations. The agreement is remarkably
good, down to ten arcminutes scales, indicating that our predictions
are robust for the Planck experiment.  Below ten arcminute scales,
where the details of the core structure dominates the power spectrum,
our analytic and simulated predictions might fail.  In the near
future, interferometer and bolometer array experiments will measure
the SZ power spectrum down to arcminutes scales, and yield
new insight into the physics of the intrahalo medium.}
\section{Analytic halo approach to the power spectrum}

\begin{figure}[t]
\begin{center}
\epsfxsize=25pc 
\epsfbox{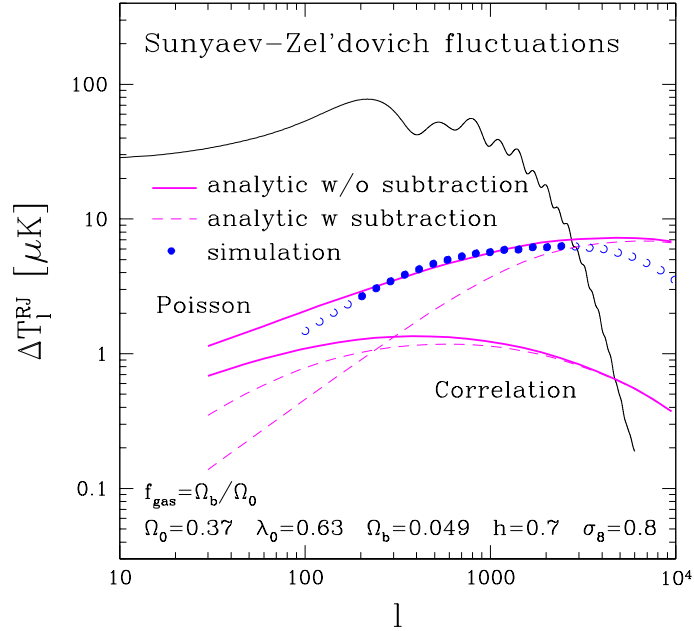} 
\end{center}
\caption{Power spectrum of SZ anisotropies from
massive halos 
($2\times 10^{12}~h^{-1}~M_\odot 
< M < 2\times 10^{15}~h^{-1}~M_\odot$)
in units of $\mu{\rm K}$ in the Rayleigh-Jeans limit.
Thick solid lines show analytic predictions
for the Poisson and the correlation terms,
while dotted lines show the predictions after subtracting 
the ROSAT X-ray flux-limited samples.
The number density of subtracted clusters is $0.8~{\rm deg^{-2}}$.
Filled and open circles show the power spectrum
measured from hydrodynamic simulations.
Filled circles correspond to the dynamic range that is 
reliably resolved in the simulation.
}
\label{fig:compare}
\end{figure}

We construct an analytic model of the power spectrum of the cosmic
microwave background (CMB) anisotropy arising from halos through the
Sunyaev-Zel'dovich (SZ) effect.  Since halos, which include clusters
of galaxies, groups, and galaxies, are discrete objects, their angular
power spectrum $C_l$ consists of a Poisson term $C_l^{(P)}$ and a
correlation term $C_l^{(C)}$ that arises from their gravitational
clustering\cite{CK88,P80}:
\begin{equation}
 \label{eq:split}
  C_l= C_l^{(P)}+C_l^{(C)}.
\end{equation}
The following describes our prescription for computing the SZ power
spectrum using the halo approach. The key ingredients in our
prescription are (1) the halo mass function $dn(M,z)/dM$, (2) the halo
bias parameter $b(M,z)$, and (3) the halo gas pressure profile
$y(\theta)$.  (1) and (2) are computed from the Press-Schechter
theory\cite{PS74,MW96,CLMP98}, while (3) is calculated assuming that
halos are virialized and isothermal. The halo approach has also been
used independently to predict the non-linear dark matter power
spectrum\cite{S00} and bispectrum\cite{MF00}. In this case, (3) is
replaced by the dark matter halo profile. These predictions are
successful in reproducing the results of N-body simulations. Thus, as
far as dark matter is concerned, our prescription is adequate to
describe the halo-halo correlation. However, once gas dynamics is
included, the degree of uncertainty in (3) increases. We will see
below that our analytic prediction nevertheless agrees well with
state-of-the-art hydrodynamic
simulations\cite{RKSP00,SBP00,SWH00,SBLT00} (see
figure~\ref{fig:compare}).

\subsection{Mass function}

One of the key ingredients in our analytic prediction is the halo mass
function $dn(M,z)/dM$, which we compute from the Press-Schechter
formalism\cite{PS74}. This formalism, which plays a central role in
our method, was used by Cole and Kaiser\cite{CK88} to calculate the
rms SZ fluctuation from halos, including both the Poisson and the
correlation terms.  Makino and Suto\cite{MS93} calculated the Poisson
term including the effect of halo profiles, assuming the $\beta$
profile.  Thanks to the recent remarkable developments of the
microwave experiments, it is now feasible to measure not only the rms
fluctuation but also the full CMB two point function, or its harmonic
transform, the angular power spectrum $C_l$, very accurately down to
arcminutes angular scales.  Atrio-Barandela and M\"ucket\cite{AM99}
calculated $C_l$ from the Poisson term of SZ fluctuations using the
Press-Schechter mass function and the $\beta$ profile.

\subsection{Bias parameter}

We predict the SZ power spectrum $C_l$ that includes the correlation
term\cite{KK99}.  For this purpose, we need the halo-halo correlation
function and its evolution with redshift. In the framework of the
Press-Schechter theory, these can be calculated from the underlying
dark matter correlation multiplied by the bias parameter
$b(M,z)$\cite{MW96,CLMP98}.  In Fourier space, this can be formulated,
to the leading order as
\begin{equation}
 \label{eq:MW96}
  P_{hh}(k,M_1,M_2,z)\simeq b(M_1,z)b(M_2,z)P_{dm}(k,z),
\end{equation}
where $P_{hh}(k)$ and $P_{dm}(k)$ are the halo-halo
and the dark matter power spectrum, respectively.  

\subsection{Halo profile}

In predicting the power spectrum of halos, we need the halo profile
which determines the small scale power.  Since the SZ effect is
produced from thermal gas pressure, we need to model both the gas
density profile and the temperature profile.  Firstly, we assume that
halos are virialized objects with an isothermal temperature
distribution.  Then, we use observationally determined halo gas
density profiles, namely the $\beta$-profile with $\beta=2/3$ and a
core radius of $0.15~h^{-1}~{\rm Mpc}$ at $M=10^{15}~h^{-1}~M_\odot$
and $z=0$.  We let the core evolve with time according to the
self-similar evolution\cite{K86}.  Note that the evolution of the core
is the {\it most uncertain} parameter in our model. Atrio-Barandela
and M\"ucket\cite{AM99} employed a different evolution model, and
found that the arcminute-scales $C_l$'s are sensitive to the core
evolution model.  We confirmed their result, and also found that the
larger angular scale spectrum is fairly insensitive to the core model.
Thus, as long as angular scales larger than about ten arcminutes are
considered, uncertainties in core evolution model do not affect the
prediction. 

\section{Power spectrum of the Sunyaev--Zel'dovich effect}

\subsection{Poisson and correlation terms}

Using the harmonic transform of the $\beta$ profile for the SZ surface
brightness distribution, $y_l(M,z)$, we get the following analytic
expressions for the angular power spectrum:
\begin{eqnarray}
  \label{eq:Cp}
  C_l{}^{(P)} &=& j_\nu^2\int dz \frac{dV}{dz}
                \int dM
                \frac{dn(M,z)}{dM} \left|y_l(M,z)\right|^2,\\
  \label{eq:Cc}
  C_l{}^{(C)} &=& j_\nu^2\int dz \frac{dV}{dz}
                       P_{dm}\left(\frac{l}{r(z)},z\right)
                       \left[\int dM
		       \frac{dn(M,z)}{dM}b(M,z) y_l(M,z)
		       \right]^2,
\end{eqnarray}
where $P$ and $C$ denote the Poisson and the correlation terms,
respectively. $j_\nu$ is the spectral function of the 
Sunyaev--Zel'dovich effect and equals $-2$ in the Rayleigh--Jeans 
regime\cite{ZS69}.
The range of the mass integral determines what objects are
considered.
We choose 
$2\times 10^{12}~h^{-1}~M_\odot < M < 2\times 10^{15}~h^{-1}~M_\odot$,
corresponding approximately to the resolution and box-size limits of our
hydrodynamic simulation\cite{RKSP00}.
The redshift integral up to $z=5$ is found to be 
sufficient for convergence.
Our choice of cosmological parameter is $\Omega_0=0.37$, $\lambda_0=0.63$,
$\Omega_b=0.049$, $h=0.7$, and $\sigma_8=0.8$.

Figure~\ref{fig:compare} shows the numerical results from equations
(\ref{eq:Cp}) and (\ref{eq:Cc}). Firstly, $C^{(C)}_l$ is always
smaller than $C^{(P)}_l$.  $C^{(P)}_l$ has a turn over at around
$l\sim 5000$, which corresponds to the typical angular size of core
radii.  Since the spectrum for $l>2000$ is very sensitive to the halo
core model\cite{AM99,KK99}, it is highly uncertain. We thus mainly focus
on larger angular scales ($l<2000$).  Note however, that the small-scale
spectrum is of great interest, as it potentially probes the
intracluster gas pressure state.  Although Planck cannot probe the
spectrum beyond $l\sim 2000$, forthcoming interferometer
and bolometer array experiments will certainly measure SZ fluctuations
on these scales.

\subsection{Subtracting nearby bright clusters}

\begin{figure}[t]
\begin{center}
\epsfxsize=25pc 
\epsfbox{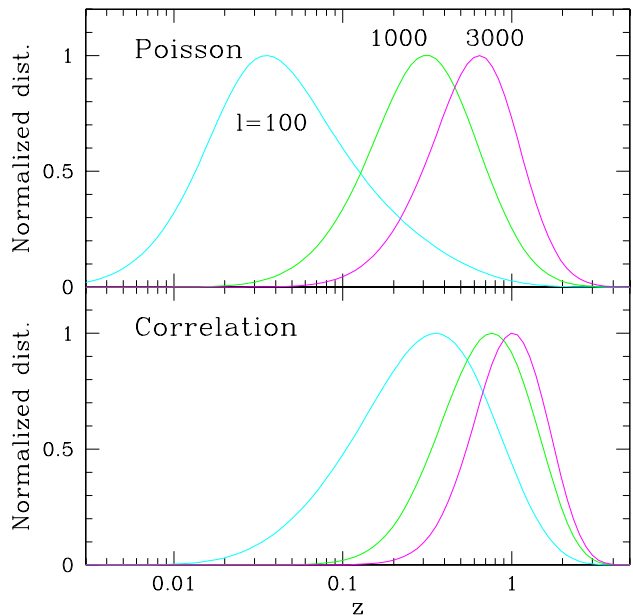} 
\end{center}
\caption{Redshift distribution  
of the SZ power spectrum in $\log z$-interval, at $l=100$, 1000, and 3000.
Amplitudes are normalized to unity.
}
\label{fig:z}
\end{figure}

Can we measure the correlation term?  Yes! 
Since nearby massive clusters (such as the Coma cluster)
dominate the Poisson term on large angular scales, the Poisson shot
noise will be substantially reduced if bright clusters are excised
from the map. Figure~\ref{fig:z} shows the redshift distribution of
$C_l$ at 3 different angular scales, $l=100$, 1000, and 3000.  Nearby
halos in $z=0.01-0.1$ dominate the Poisson $C^{(P)}_{100}$, while the
correlation $C^{(C)}_{100}$ come from $z=0.1-1$.  This clear
separation in redshift space allows us to remove the Poisson shot
noise, while preserving the correlation term.

One way of doing this is to subtract a homogeneous sample of local
clusters from the SZ map. Figure~\ref{fig:compare} shows the SZ power
spectrum before (solid lines) and after (dashed lines) subtracting an
X-ray selected cluster sample with the flux limit $>10^{-13}~{\rm
erg~cm^{-2}~s^{-1}}$ in the ROSAT ($0.5-2$ keV) band. The Poisson
spectrum is greatly reduced, while the correlation spectrum is less
affected by the subtraction. The correlation term actually dominates
over the Poisson term for $l<200$.  The number density of subtracted
clusters is $0.8~{\rm deg^{-2}}$. For this figure, the X-ray flux was
calculated by the method of Kitayama and
Suto\cite{KS97}. Alternatively, one can identify bright SZ clusters in
the CMB map itself, and then subtract them\cite{KK99}.  For example,
$0.9~{\rm deg^{-2}}$ clusters can be removed for ones brighter than
$50~{\rm mJy}$ at 350~GHz.

\begin{figure}[t]
\begin{center}
\epsfxsize=25pc 
\epsfbox{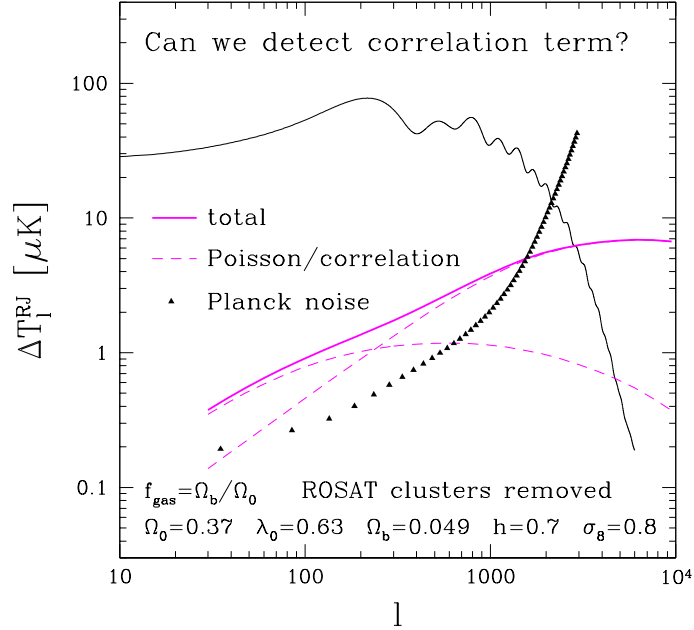} 
\end{center}
\caption{Comparison of the predicted signal from the correlation term
to the expected sensitivity of the Planck experiment 
(filled triangles) to the Sunyaev-Zel'dovich effect.
The noise power spectrum has been binned into bands with 
$\Delta l=50$.
The thick solid line shows the total spectrum, while
the dashed lines show the Poisson and the correlation terms separately.
The ROSAT X-ray flux-limited sample was removed.
The number density of subtracted clusters is $0.8~{\rm deg^{-2}}$.
Note that the plotted sensitivity is the residual noise level after
subtracting the primary CMB signal from the map, on the basis of a 
multi-frequency analysis.
}
\label{fig:noise}
\end{figure}

\subsection{Can we detect the correlation term?}

Even if the correlation term overcomes the Poisson term, 
the primary CMB fluctuations may prevent us from measuring
the correlation term on degree scales.
Can we really measure the correlation term? Yes! 
Thanks to the unique shape of the frequency spectrum $j_\nu$
of the SZ effect\cite{ZS69}, we can subtract 
the primary CMB, and hopefully other foregrounds from 
the map. In other words, we expect to be able to derive a
{\it cleaned} SZ map.
Cooray, Hu and Tegmark\cite{CHT00} demonstrated
this technique:
using the frequency separation, 
they obtained the expected noise level for the SZ power spectrum
after subtracting the primary CMB and other modeled sources of foreground.
Figure~\ref{fig:noise} compares the signal of the correlation term
to the resulting SZ noise power spectrum, binned into bands with 
$\Delta l=50$.
The result is so encouraging that we can hope to 
measure the contribution from the correlation term fairly accurately
at $l=30-200$.
Since the degree-scale spectrum is insensitive to 
the highly uncertain core structures of halos,
our predictions are robust on these scales.
Thus, measuring the correlation term cleanly probes
the clustering of distant halos. Such a measurement
has not been achieved yet, mainly because of the lack 
of sensitivity of optical and X-ray
surveys to such distant clusters and groups.
The fact that the SZ effect is very sensitive to high redshift,
is a consequence of the known fact that the SZ surface 
brightness is independent of $z$.

\section{Comparison to the moving-mesh hydrodynamic simulation}

\begin{figure}[t]
\begin{center}
\epsfxsize=25pc 
\epsfbox{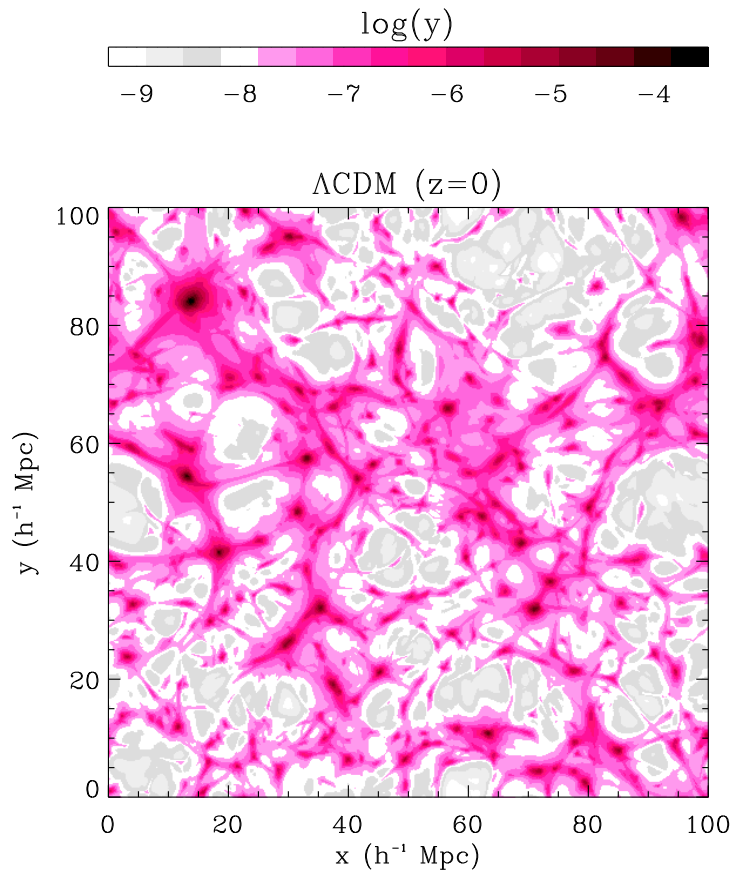} 
\end{center}
\caption{Snap shot of $y$-parameter in the moving-mesh 
hydrodynamic simulation at $z=0$.
The box size is $L=100~h^{-1}~{\rm Mpc}$, and the number of cells is 
$N=128^3$. The effective spatial resolution at $z=0$ is 
$\sim 80~h^{-1}~{\rm kpc}$.
Our choice of cosmological parameters is $\Omega_0=0.37$, $\lambda_0=0.63$,
$\Omega_b=0.049$, $h=0.7$, and $\sigma_8=0.8$.
The angular power spectrum of the SZ effect from the
simulation is plotted in 
figure~\ref{fig:compare}. For the simulation, the mean 
density-weighted temperature
at present is 0.25~keV, and the mean $y$-parameter is 
$1.7\times 10^{-6}$, in good agreement with the Press-Schechter
predictions.
}
\label{fig:ymap}
\end{figure}

We carried out adiabatic hydrodynamic simulations\cite{RKSP00} 
using the moving-mesh
hydrodynamic code written by Pen\cite{P9598}.
The code captures the advantages of both the Lagrangian particle based and 
the Eulerian grid based codes by allowing the grid meshes to deform 
along potential flow lines.
This strategy makes it possible to increase the resolution by twenty
fold over previous Cartesian grid Eulerian schemes, at a low computational 
cost. 

Figure~\ref{fig:ymap} shows 
a snap shot of the $y$-parameter in the simulation at $z=0$.
The box size is $L=100~h^{-1}~{\rm Mpc}$, and the number of cells is 
$N=128^3$. Since the linear compression factor is about 10, 
the effective spatial resolution at $z=0$ is $\sim 80~h^{-1}~{\rm kpc}$.
Our cosmological parameters are $\Omega_0=0.37$, $\lambda_0=0.63$,
$\Omega_b=0.049$, $h=0.7$, and $\sigma_8=0.8$.
The mass range resolved in this simulation is
$2\times 10^{12}~h^{-1}~M_\odot < M < 2\times 10^{15}~h^{-1}~M_\odot$.

Figure~\ref{fig:compare} compares the angular power spectrum 
of the SZ effect from the simulation (filled and open circles)
to the analytic predictions (solid lines).
The filled circles highlight the $l$-range, $200<l<2000$, 
which is reliably described in the simulation.
The agreement between the simulation and the analytic prediction is
impressively good.
Likewise, the mean density weighted temperature
at present is 0.25~keV, and the mean $y$-parameter is 
$1.7\times 10^{-6}$, in good agreement with the Press-Schechter
predictions.
Thus, our halo approach to the SZ effect works as long as 
relatively massive halos resolved in this simulation 
are considered. We conclude that the SZ power spectrum 
in $200<l<2000$ is dominated by these massive halos.

\section{Discussion}

Our analytic predictions agree well with the adiabatic 
hydrodynamic simulations; however, our halo gas pressure model 
disagrees somewhat with observations.
This disagreement affect our analytic predictions and 
simulated results qualitatively
on smaller angular scales, say, $l>2000$.
For example, 
the self-similar model appears to have difficulties in explaining
the X-ray luminosity--temperature relation\cite{AE99},
the mass--temperature relation\cite{FRB00}, and 
the central entropy--temperature relation\cite{PCN99}.
The departure from the adiabaticity due to so-called preheating is 
thought to be a solution to these problems. 
If this is so, the adiabatic simulation would not describe the intrahalo gas
state accurately.
The resolution of these disagreements seems to require larger core 
radii than those predicted by the self-similar model. 
This would tend to suppress the small scale
power spectrum, as compared to the adiabatic case\cite{SWH00}.
Including those effects in analytic predictions is rather 
challenging, but is important. Interferometer and 
bolometer array CMB experiments are expected to measure the 
SZ power spectrum accurately down to arcminutes scales, in the near future.
These experiments will thus certainly further our understanding
of the intracluster medium.

\section*{Acknowledgments}

We would like to thank Wayne Hu for providing the noise power spectrum.
E. K. would like to thank Uro$\check{\rm s}$ Seljak for 
frequent discussions.
E. K. and T. K. acknowledge fellowships
from the Japan Society for the Promotion of Science.
D. N. S. is partially supported by the MAP/MIDEX program.
A. R. is supported by an EEC TMR grant.
Computing support from the National Center for Supercomputing
Applications is acknowledged.


\end{document}